\documentclass[12pt]{article}
\pdfoutput=1
\usepackage{amsmath,amssymb,epsfig,amsfonts,graphicx,euscript}
\usepackage[pdftex,bookmarks,bookmarksnumbered,linktocpage,pdfstartview=FitH]{hyperref}
\hypersetup{colorlinks,%
citecolor=red,%
filecolor=blue,%
linkcolor=blue,%
urlcolor=blue,%
pdftex}
\usepackage[all]{hypcap}
\numberwithin{equation}{section}
\usepackage{cite}
\newcommand{\bse}{\begin{subequations}}
\newcommand{\ese}{\end{subequations}}
\newcommand{\be}{\begin{equation}}
\newcommand{\ee}{\end{equation}}
\newcommand{\bea}{\begin{eqnarray}}
\newcommand{\eea}{\end{eqnarray}}
\newcommand{\ba}{\begin{array}}
\newcommand{\ea}{\end{array}}

\begin{document}
\hfill%
\vbox{
    \halign{#\hfil        \cr
           IPM/P-2017/019\cr
                     }
      }
\vspace{0.5cm}
\begin{center}
{\Large{\textbf{A Classical String in Lifshitz-Vaidya Geometry}}} \\
\vspace*{1cm}
\begin{center}
{\bf A. Hajilou\footnote{$a_{-}$hajilou@sbu.ac.ir}$^{,a}$, M. Ali-Akbari\footnote{{\rm{m}}$_{-}$aliakbari@sbu.ac.ir}$^{,a}$, F. Charmchi\footnote{charmchi@ipm.ir}$^{,b}$}\\%
\vspace*{0.4cm}
{\it {${}^a$Department of Physics, Shahid Beheshti University G.C., Evin, Tehran 19839, Iran}}  \\
{\it {${}^b$School of Particles and Accelerators, Institute for Research in Fundamental Sciences (IPM),
P.O.Box 19395-5531, Tehran, Iran}}  \\

\vspace*{0.5cm}
\end{center}
\end{center}

\vspace{.2cm}
\bigskip
\begin{center}
\textbf{Abstract}
\end{center}
We study the time evolution of the expectation value of a rectangular Wilson loop in strongly anisotropic time-dependent plasma using gauge-gravity duality. The corresponding gravity theory is given by describing time evolution of a classical string in the Lifshitz-Vaidya background. We show that the expectation value of the Wilson loop oscillates about the value of the static potential with the same parameters after the energy injection is over. We discuss how the amplitude and frequency of the oscillation depend on the parameters of the theory. In particular, by raising the anisotropy parameter, we observe that the amplitude and frequency of the oscillation increase.


\newpage

\tableofcontents
\section{Introduction and Result}
Quark-gluon plasma, as a new phase of matter, is produced at Relativistic Heavy Ion Collider (RHIC) and Large Hadron Collider (LHC) by colliding two pancakes of heavy nuclei such as Gold(Au) or lead(Pb) at a relativistic speed \cite{CasalderreySolana:2011us}. Through viscous hydrodynamical simulations, it is realized that viscosity over entropy density is small, i.e. $\eta/s=1/4\pi$ \cite{Shuryak:2004cy, Shuryak:2003xe}, and it is then a strong indication that the plasma is strongly coupled. Furthermore, at early times, the plasma is far from equilibrium and after a certain time viscous hydrodynamic description can be applied. Although the viscous hydrodynamics is applicable during most of the time evolution, a significantly different pressure between longitudinal and transverse directions exists indicating that the plasma created in the heavy ion collision is anisotropic.

Since the plasma is strongly coupled, it is not reliable to describe various properties of the plasma by applying perturbation method. As a results, as a non-perturbative method, gauge-gravity duality provide a novel approach for studying the strong coupling limit of a large class of non-abelian quantum gauge theories \cite{Witten:1998qj}. According to this duality, a strongly coupled gauge theory defined in a $d$-dimensional space-time corresponds to a classical gravity in a $d+1$-dimensional space-time \cite{Witten:1998qj, Maldacena:1997re, CasalderreySolana:2011us}. Therefore, different questions in the strongly coupled gauge theory can be translated into corresponding problems in the classical gravity. This duality has been frequently applied to study various aspects of the strongly coupled systems such as static potential energy between a quark and anti-quark pair \cite{Maldacena:1998im}, jet quenching parameter \cite{Liu:2006ug}, thermalization and isotropization process \cite{Heller:2013oxa, Ali-Akbari:2016sms}. For more details see \cite{CasalderreySolana:2011us} and references therein.

Finding static potential energy between a quark and anti-quark pair, or equivalently quark-anti-quark bound state, living in the plasma is an interesting problem that has been attracted a lot of attention. This problem has been firstly addressed in \cite{Maldacena:1998im} and then its generalization has been widely discussed in the literature. Concisely, in order to calculate the static potential energy between the pair we need to compute expectation value of a rectangular Wilson loop in the strongly coupled plasma. The holographic dual of the rectangular Wilson loop is given by a classical string suspended from two points (corresponding to quark and anti-quark), hanging down in extra dimension with appropriate boundary conditions. Using this idea, static potential energy is studied in different gauge theories with holographic duals and it recently generalizes to a time-dependent case in \cite{Ali-Akbari:2015ooa}.  In the time-dependent case, the time evolution of the expectation value of the Wilson loop during the energy injection into the gauge theory is investigated. Holographic-ally, the mentioned system corresponds to the time evolution of the classical string in the AdS-Vaidya background. As a toy model,  the AdS-Vaidya background is dual to thermalization process in the gauge theory \cite{Balasubramanian:2011ur}. 

The Lifshitz-like background, which is holographic-ally dual to an anisotropic plasma, is applied to investigate different properties of the anisotropic plasma. One of the things that makes the Lifshitz-like background interesting is that holographic estimates of the total multiplicity can fit the experimental data at high energy for certain values of critical exponent \cite{Arefeva:2014vjl}.  It was also shown that the Lifshitz-like background can be considered as the IR limit of the 10-dimensional IIb supergravity anisotropic background suggested in \cite{Mateos:2011tv}. Vaidya solutions in the Lifshitz-like background have been found in \cite{Arefeva:2016phb}. 

In present work we calculate the time evolution of the expectation value of the rectangular Wilson loop in the Lifshitz-Vaidya background\footnote{Note that this background can be considered as the IR limit of a 10-dimensional solution of IIb supergravity similar to the case suggested in \cite{Mateos:2011tv}. }. More precisely, we would like to discuss the effect of various parameters on the time evolution of the Wilson loop. In particular, the critical exponent $\nu$, or equivalently anisotropy parameter, plays an important role in our work.
Our main findings can be summarized as follows:
\begin{itemize}
\item The quark-anti-quark bound state is excited in the anisotropic plasma due to the energy injection. The characteristic of the excited bound state \textit{does depend} on anisotropy parameter which, in our case, is given by the critical exponent of the Lifshitz-like metric. In fact, for larger values of the anisotropy parameter, when the other parameters have been fixed, the excited bound state oscillates with larger oscillation frequency $f$ and amplitude $A$.
\item To compare with the real plasma produced at RHIC or LHC, the case of $\nu=4$ is more reasonable \cite{Arefeva:2014vjl}. Our numerical results show that $\frac{f_{\nu=4}}{f_{\nu=1}}\simeq 8$ and $\frac{A_{\nu=4}}{A_{\nu=1}}\simeq 2.1$ for the same values of the transition time, final temperature and distance $l$ between quark and anti-quark. By transition time we mean how slow or fast the energy has been injected into the system under study. As a matter of fact, the anisotropy of the system substantially influences the bound state living in the plasma. 
\item We observe that the oscillation frequency of the excited bound state \textit{depends} on the transition time. In other words, for fast (slow) energy injection the bound state is excited with larger (smaller) oscillation frequency for a fixed value of anisotropy parameter at fixed temperature. Larger values of the anisotropy parameter have larger oscillation frequencies. Similarly, the amplitude of the oscillation increases for smaller transition time $k$.
\item Our numerical calculations show that the final temperature and the oscillation frequency are \textit{independent}. It happens for all cases with or without anisotropy parameter.  At fixed temperature, we observe that the anisotropy parameter, the amplitude and frequency of oscillation increase together. 
However, for given anisotropy parameter, the amplitude of the oscillation and temperature increase together while the oscillation frequency does not change.
%
\item Another result is that the frequency and amplitude of the oscillation depend on the distance between quark and anti-quark. By raising the distance, both frequency and amplitude increase.
\end{itemize}

\section{Review on the static and time-dependent backgrounds}
In this section, we will give a brief review on the background used to calculate (time-dependent) expectation value of the Wilson-loop.  
The non-zero temperature Lifshitz-like metric is given by
\be\label{BHlif} %
 ds^2=\frac{1}{z^2}(- f(z) dt^2 + d{x}_1^2) + \frac{1}{z^\frac{2}{\nu}}(d{x}_2^2+ d{x}_3^2) + \frac{dz^2}{z^2f(z)},
\ee %
where
\be %
\label{lifshitzmass}
f(z)=1 - m z^{\frac{2}{\nu}+2},
\ee %
and $\nu$ is the critical exponent or Lifshitz parameter. As is clearly seen, there is an anisotropy between $x_1$ and other spatial coordinates, i.e. $x_2$ and $x_3$. According to gauge-gravity duality, this anisotropy in the gravity is identified with an isotropy on the gauge theory side or, in other words, the gauge theory live on the anisotropic background.  The horizon is located at $z_h=m^{-\nu/(2+2\nu)}$  and therefore the Hawking temperature, corresponding to the temperature of the gauge theory, is given by $T=\frac{1}{\pi z_h}(\frac{\nu+1}{2\nu})$. The boundary lies at $z=0$ and the metric approaches Lifshitz-like geometry asymptotically. In this metric when we put $\nu=1$, the metric reduces to AdS-BH space-time metric. This background has been extensively applied in the literature to discuss various aspects of the anisotropic plasma, for instance see \cite{Ageev:2016gtl}.

A generalization of the above background to the case of Lifshitz-Vaidya metric is given by \cite{Arefeva:2016phb} 
\be\label{vlif}%
 ds^2=\frac{1}{z^2}(- F(\bar{v},z) d\bar{v}^2 -2d\bar{v}dz+ d{x}_1^2) + \frac{1}{z^{\frac{2}{\nu}}} (d{x}_2^2+ d{x}_3^2),
\ee %
where 
\be\label{MV}\begin{split} %
F(\bar{v},z)=1 - M(\bar{v})z^{\frac{2}{\nu}+2},\cr
T(\bar{v})=(\frac{\nu+1}{2\pi\nu})M(\bar{v})^{\frac{\nu}{2+2\nu}}.
\end{split}\ee %
The arbitrary function $M(\bar{v})$, related to the temperature of the gauge theory, represents the mass of the black hole which changes as time passes by until it reaches a constant value.
The above metric is written in the Eddington-Finkelstein coordinates where the radial direction is represented by $z$. The coordinate $\bar{v}$ shows the null direction where, at the boundary, $\bar{v}$ is equal to the time coordinate of the
gaue theory, i.e. $t = \bar{v} |_{z=0}$. Note that, in the case of the $\nu=1$, the above metric reduces to AdS-Vaidya metric.

According to gauge-gravity duality, the Lifshitz-Vaidya metric on the gravity side resembles thermalization process in the anisotropic gauge theory.
Various types of energy injection are identified by the form of the
functions chosen for the time-dependent function $M(\bar{v})$. These different forms have been investigated and classified in \cite{Amiri-Sharifi:2016uso} and it seems that the final qualitative results are independent of the form of the functions. Here, temperature is turned on exactly at $\bar{v} = 0$ and reaches its exact final maximum value, $T_f$ at some finite time. Therefore, the functions for $M(\bar{v})$ that we will work with can be considered as
\bea %
 M(\bar{v})= M_f \left\{%
\begin{array}{ll}
    0 & \bar{v}<0, \\
    k^{-1}\left[\bar{v}-\frac{k}{2\pi}\sin(\frac{2\pi \bar{v}}{k})\right] & 0 \leqslant \bar{v} \leqslant k, \\
    1 & \bar{v}>k ,\\
\end{array}%
\right.
\eea %
where the transition time $k$ is the time interval in which the mass of the black hole increases from zero to $M_f$ which is constant. Note that the radius of the event horizon is $z_h=M_f^{-\frac{\nu}{2+2\nu}}$ and therefore $T_f=\frac{1}{\pi z_h}(\frac{\nu+1}{2\nu})$. 

The transition time $k$ plays a central role in energy injection into the system. As a matter of fact, for small values of $k$ a universal behaviour is observed \cite{Ali-Akbari:2015gba}. By universal behaviour we mean the re-scaled equilibration time $k^{-1}t_{eq}$ is independent of the final value of the temperature. This result is perhaps common to all strongly coupled gauge theory with gravity dual. For this reason we are particularly interested in studying two different regimes for the transition time, that is $k<1$ and $k>1$. Thus, in the following we choose $k=0.3$ and 3.

\section{Probe classical string}
In this section, using gauge-gravity duality, we will obtain the static potential energy between a quark and an anti-quark (or equivalently expectation value of the Wilson loop) in the anisotropic plasma describing by the background \eqref{BHlif}. Then the expectation value of the Wilson loop is numerically computed in the Lifshitz-Vadiya metric \eqref{vlif}. In fact, time-dependent solution oscillates about the static one found in the background \eqref{BHlif}.

The dynamics of a classical string in an arbitrary background is described by 
\be %
\label{action}
 S=\frac{-1}{2 \pi \alpha'} \int d\tau d\sigma  \sqrt{- \det(g_{ab})}.
\ee %
where $g_{ab}$ is the induced metric on the world-sheet and is defined by $g_{ab}=G_{MN}\frac{\partial X^M}{\partial \xi^a} \frac{\partial X^N}{\partial \xi^b}$. $X^M$ ($\xi^a=\tau,\ \sigma$) denotes the space-time (world-sheet) coordinates and $G_{MN}$ is the background metric. In the gauge theory, the static potential energy between a quark and an anti-quark is evaluated by using the expectation value of the Wilson loop on a rectangular loop, ${\cal{C}}$, that contains two sides, time $\cal{T}$ and distance $l$, where the length of time direction is much larger than the distance between the quarks , i.e. ${\cal{T}}\gg l$. Therefore, we finally have \cite{CasalderreySolana:2011us}
\be\label{staticwilson}
\langle W({\cal{C}}) \rangle  =e^{-i(2 m + V(l)){\cal{T}}},
\ee%
where $m=\frac{\sqrt{\lambda}}{2\pi}\int_{\epsilon}^{z_h}\,\frac{dz}{z^2}$ is the rest mass of the quarks and $V(l)$ is the potential energy between them. $\epsilon$ is IR regulator in the bulk and from UV/IR connection, its correspondence is UV cut-off in the boundary theory. On the other hand, according to the gauge-gravity dictionary, the expectation value of the Wilson loop is dual to the on-shell action of a string that its endpoints is separated by a distance $l$ \cite{CasalderreySolana:2011us}. Thus 
\be\label{wilson}
\langle W({\cal{C}})\rangle =e^{i S({\cal{C}})}.
\ee%
We will now proceed to calculate $S({\cal{C}})$ for the rectangular loop ${\cal{C}}$.

\subsection{On the static Lifshitz black hole background}
In the static case, we work with the metric \eqref{BHlif}. It is then convenient to choose the world-sheet coordinates as
\be
\tau= t ,\ \sigma=x_3\equiv x,
\ee
and following ansatz for the other coordinates
\be %
z=z(x),\ \ x_1={\rm{constant}},\ \ x_2=\rm{constant}.
\ee
Note that since we consider ${\cal{T}}\gg l$, one can assume that the world sheet is translationally invariant along the $\tau$-direction. Therefore, the string action \eqref{action} leads to 
\be\label{static-action}
S=\frac{-1}{2\pi\alpha'}\int_{-\frac{l}{2}}^{\frac{l}{2}}\,dt dx\frac{1}{z^2}
\sqrt{{z^\prime}^2+z^{2-\frac{2}{\nu}}\,{f}(z)},
\ee
where $z'\!=\!dz/dx$. 
Since the Lagrangian does not explicitly depend on $x$, we can use the associated Hamiltonian, which is the constant of motion, to obtain the static solution.
As a result we get
\be\label{variation}
z'(x)=\pm\frac{{z_\ast}^{1+\frac{1}{\nu}}~f(z)}{\sqrt{{f}(z_\ast)}~z^{\frac{2}{\nu}}}~
\sqrt{1-\frac{{f}(z_\ast)}{{f}(z)}\left(\frac{z}{{z_\ast}}\right)^{2+\frac{2}{\nu}}},
\ee
where $z'(x)\!=\!0$ at $z\!=\!z_*$ where $z_*$ is the turning point of the string. The above equation allows us to express $x$ as a function of $z$. 
Using the explicit form of ${f}(z)$ and new coordinate $y=z_*/z$, one gets
\be\label{initial}
\int_{\pm\frac{l}{2}}^x\,dx=\mp z_{\ast}^{\frac{1}{\nu}}\sqrt{1-y_h^{2+\frac{2}{\nu}}}
\int_1^y\frac{dy}{\sqrt{\left(y^{2+\frac{2}{\nu}}-1\right)\left(y^{2+\frac{2}{\nu}}-y_h^{2+\frac{2}{\nu}}\right)}},
\ee
where $y_h=z_*/z_h$. The potential energy between the quark and anti-quark can be obtained as explained in 
\cite{Maldacena:1998im}. 
This energy is divergent due to the infinite mass of the quarks.
We should subtract the action of two strings stretched between the boundary $(z\!=\!0)$ and the horizon $(z\!=\!z_h)$
 from the action (\ref{static-action}) to obtain a finite and regular result, which read 
\begin{align}
V_{\textrm{static}}&=
\frac{-1}{{\cal T}}\left[S^{\textrm{on-shell}}
+\frac{{\cal T}}{\pi\alpha^\prime}
\left(\int_\epsilon^{z_*}+\int_{z_*}^{z_h}\right)\frac{dz}{z^2}\right]\nonumber \\
&=\frac{1}{\pi\alpha^\prime}\left[\frac{1}{z_\ast}\int_1^\infty\,dy
\left(\sqrt{\frac{y^{2+\frac{2}{\nu}}-y_h^{2+\frac{2}{\nu}}}{y^{2+\frac{2}{\nu}}-1}}-1\right)
-\left(\frac{1}{z_{\ast}}-\frac{1}{z_h}\right)\right].
\end{align}
The on-shell action can be obtained by replacing \,\eqref{variation} into \,\eqref{static-action}.
In the pure Lifshitz background, that is $z_h\to \infty$, \,\eqref{initial} reduces to
\be
x=\pm\frac{l}{2}\mp\frac{z^{1+\frac{2}{\nu}}}{(1+\frac{2}{\nu})z_\ast^{1+\frac{1}{\nu}}}\,
{}_2F_1\left(\frac{1}{2},\frac{2+\nu}{2+2\nu},\frac{4+3\nu}{2+2\nu},\frac{z^{2+\frac{2}{\nu}}}{z_\ast^{2+\frac{2}{\nu}}}\right).
\ee
We use this result as the initial condition for the dynamical background case.

\subsection{In the Lifshitz-Vaidya background}
In the static case, as it was stated, in order to find potential energy the condition ${\cal{T}}\gg l$ is essential.  Unfortunately, in the Vaidya-Lifshitz background, this condition does not valid anymore and therefore we have to calculate time-dependent Wilson loop. In other words, instead of \eqref{staticwilson}, we have 
\be\label{timewilson}
\langle W({\cal{C}})\rangle =e^{-i \int  dt\  {\cal{W}}(l,t)}.
\ee%
The gauge-gravity correspondence proposes that ${\cal{W}}(l,t)$ is dual to the string on-shell action in which the time coordinate is not integrated over. Note that the above equation clearly reduces to \eqref{staticwilson} in the time-independent background. Due to UV cut-off ($z\rightarrow 0$) an infinity appears in ${\cal{W}}(l,t)$. As a result, similar to static case, we regularize \eqref{timewilson} as follows
\be\begin{split}%
{\cal{W}}_R(l,t) &= {\cal{W}}(l,t) - 2 m \cr
&= \int  d\sigma \left(\sqrt{- \det (g_{ab})}\right)_{\rm{on-shell}} - 2 m.
\end{split}\ee%
In order to compute string on-shell action we use the numerical method introduced in \cite{Ali-Akbari:2015ooa, Ishii:2014paa}. To do so, we use the null coordinates $(u,v)$ on the string world-sheet. Thus all background coordinates  on the world-sheet depend on the null coordinates and apart from $\bar{v}=V(u,v)$, $z=Z(u,v)$ and $x_3=X(u,v)$, we set two other coordinates to zero. Substituting this ansatz in the action \eqref{action}, it is easy but lengthy to find the following equations of motion 
\be\label{eom}\begin{split}%
V_{,uv}&=
\left( \frac{F_{,Z} }{2} -\frac{F}{Z}\right) V_{,u} V_{,v} + \frac{1}{\nu Z^{\frac{2}{\nu}-1}} X_{,u} X_{,v}\, ,
\cr
Z_{,uv}&=\left(\frac{F^2}{Z} - \frac{F }{2} F_{,Z} -\frac{1}{2} F_{,V}\right) V_{,u} V_{,v} 
+ \left(\frac{F}{Z} -\frac{F_{,Z}}{2}\right) \left(Z_{,u} V_{,v}+Z_{,v} V_{,u}\right)\cr
&+ \frac{2}{Z} Z_{,u} Z_{,v}   - \frac{F}{\nu Z^{\frac{2}{\nu}-1}} X_{,u} X_{,v}\, ,\cr
X_{,uv}& = \frac{Z_{,u} X_{,v} + Z_{,v} X_{,u}}{\nu Z}.
\
\end{split}\ee%
Since $u$ and $v$ are null coordinate on the world-sheet, we have to impose two constraint equations 
\be\label{cons}\begin{split}%
C_1 = \frac{1}{Z^2} (F(V,Z) V_{,u}^2 + 2 V_{,u}  Z_{,u} -Z^{2-\frac{2}{\nu}} X_{,u}^2) & = 0,\cr
C_2 = \frac{1}{Z^2} (F(V,Z) V_{,v}^2 + 2 V_{,v}  Z_{,v} - Z^{2-\frac{2}{\nu}} X_{,v}^2 ) & = 0,
\end{split}\ee%
corresponding to $g_{uu}=0$ and $g_{vv}=0$, respectively. 

In order to solve the equations of motion \eqref{eom} subject to constraint equations \eqref{cons}, we need to specify the initial and boundary conditions. These conditions are similar to the ones considered in \cite{Ishii:2014paa}. Here we do not repeat the details of calculations and only state the final results. Therefore, we have
\begin{itemize}
\item \textbf{Boundary condition:}\\
Based on the discussions in \cite{Ishii:2014paa}, by fixing the diffeomorphism on the string world-sheet
one may choose the AdS boundary to be at $u = v$
for one of the endpoints and $u =v+L$ for the other one.
Then, on the AdS boundary, the appropriate boundary conditions on $Z$ and $X$ are 
\be\begin{split}%
Z|_{u=v}&=0~;~~~~X|_{u=v}=\frac{-l}{2},\cr
Z|_{u=v+L}&=0~;~~~~X|_{u=v+L}=\frac{l}{2}.
\end{split}\ee
One can find the rest of the boundary conditions by expanding the fields near the boundary $u=v$ as follows
\bea%
V(u,v)&=&V_0(v) + V_1(v) (u-v) + ...~,~~~~~~~~\\  
Z(u,v)&=&Z_1(v) (u-v) +Z_2(v) (u-v)^2+ ...~,~~~~~~~~\\
X(u,v)&=&\frac{-l}{2} + X_1(v) (u-v) + ...~,~~~~~~~~
\eea%
Then demanding the regularity condition at $u\!=\!v$ and $u\!=\!v+L$ the rest of the boundary condition can be found. Furthermore, the consistency of the results with the constraint equations \eqref{cons} must be checked. The final results for boundary conditions in terms of anisotropic parameter $\nu$ at $u=v$ are:
\bse\begin{align}%
\label{bc1}
V(u,v)&=V_0(v) + {\cal{O}}\left((u-v)^m\right),\\ \nonumber
\label{bc2}  
Z(u,v)&=\frac{\dot{V}_0(v)}{2} (u-v) + \frac{\ddot{V}_0(v)}{4} (u-v)^2\\
&+\frac{\dddot{V}_0(v)}{12} (u-v)^3 + {\cal{O}}\left((u-v)^n\right),\\ 
X(u,v)&=\frac{-l}{2} + {\cal{O}}\left((u-v)^r\right), 
\end{align}\ese%
\begin{table}[ht]
\caption{Appropriate numbers for boundary conditions}
\vspace{3 mm}
\centering
\begin{tabular}{c c c c c}
\hline\hline
~~$ \nu$ ~~   &   ~~ $1$ ~~ & ~~ $2$ ~~ & ~~ $3$ ~~ & ~~ $4$ ~~\\[0.5ex]

\hline
$m$ & 5 & 4 & 6 & 8\\
$n$ & 4 &  4  & 4 & 4\\
$r$ & 3 &  2 & 6 & 7\\
\hline
\end{tabular}\\[1ex]
\label{list}
\end{table}

where $\dot{V}(v)=\frac{dV(v)}{dv}$ and so on. $m,n$ and $r$ are listed in table \ref{list}. The above equations then imply that
\be%
Z_{,uv}|_{u=v} = 0,\ 2 Z_{,u}|_{u=v} = \dot{V}_0(v).
\ee%
One can easily check that for the another boundary $u=v+L$ the results are the same. We refer interested reader to \cite{Ishii:2014paa} for more details. 

\item \textbf{Initial condition:}\\
To obtain the initial conditions for the variables $V$, $Z$ and $X$ we use the constraint equations and the static solution \eqref{variation}. Notice that in this equation we replace $z$ and $x$ with the capital ones and $f(z)=1$. Since $V_{,v}>0$ at the boundary, therefore by using the boundary conditions \eqref{bc1} and \eqref{bc2}, $Z_{,u}>0$ and $Z_{,v}<0$. Applying these conditions on $Z$ and $V$ derivatives and  using $X_{,u}|_{Z=0}=X_{,v}|_{Z=0}=0$, the constraint equations \eqref{cons} lead to
\bea%
\label{eqV1}
V_{,u}
&= Z_{,u} \bigg(-1+\sqrt{1+Z^{2-\frac{2}{\nu}}\big(\frac{dX}{dZ}\big)^2}\bigg) ,~\\
\label{eqV2}
V_{,v}
&= Z_{,v} \bigg(-1-\sqrt{1+Z^{2-\frac{2}{\nu}}\big(\frac{dX}{dZ}\big)^2}\bigg).
\eea%
By taking the derivative of the equation \eqref{eqV1} with respect to $v$ and of the equation \eqref{eqV2} with respect to $u$, we obtain
\be%
\label{master}
Z_{,uv} \bigg( \sqrt{1+Z^{2-\frac{2}{\nu}}\big(\frac{dX}{dZ}\big)^2}\bigg) + Z_{,v} Z_{,u} \bigg(\sqrt{1+Z^{2-\frac{2}{\nu}}\big(\frac{dX}{dZ}\big)^2}\bigg)_{,Z}=0,
\ee%
and it can be then written as
\be\label{equ}%
\bigg( Z_{,u} \sqrt{1+Z^{2-\frac{2}{\nu}}\big(\frac{dX}{dZ}\big)^2}  \bigg)_{,v}=0.
\ee%
One can substitute $\frac{dX}{dZ}$ into \eqref{equ} by using \eqref{variation} and we then have 
\be%
\label{solZ}
Z \, _2F_1\bigg(\frac{1}{2},\frac{\nu}{2+2\nu};\frac{2+3\nu}{2+2\nu};\frac{Z^{2+\frac{2}{\nu}}}{Z_*^{2+\frac{2}{\nu}}}\bigg) = \phi(u) - \phi(v),
\ee%
where $\phi(y)$ is an arbitrary function. The form of the right hand side of the above equation is fixed by applying the fact that the left hand side is zero at $u=v$. 

By integrating \eqref{variation}, we get the initial configuration for $X(u,v)$ as follows
\be%
X(u,v) = \frac{l}{2} - \frac{Z^{1+\frac{2}{\nu}}}{(1+\frac{2}{\nu}) Z_*^{1+\frac{1}{\nu}}} \, _2F_1\left(\frac{1}{2},\frac{2+\nu}{2+2\nu};\frac{4+3\nu}{2+2\nu};\frac{Z^{2+\frac{2}{\nu}}}{Z_*^{2+\frac{2}{\nu}}}\right),
\ee%
where $Z_*$ is the turning point of the string. Since $X(u,v)=0$ at $Z=Z_*$, we have 
\be%
Z_*=\left(\frac{2+\nu}{2\nu}\frac{l}{ _2F_1\left(\frac{1}{2},\frac{2+\nu}{2+2\nu};\frac{4+3\nu}{2+2\nu};1\right)}\right)^{\nu}.
\ee%

Also, the initial condition on $V(u,v)$ is obtained from \eqref{eqV1} and \eqref{eqV2}
\bea%
V(u,v)&=- Z \left(1-\, _2F_1\left(\frac{1}{2},\frac{\nu}{2+2\nu};\frac{2+3\nu}{2+2\nu};\frac{Z^{2+\frac{2}{\nu}}}{Z_*^{2+\frac{2}{\nu}}}\right)\right)+\chi(v),~~~~~~~\\
V(u,v)&=- Z \left(1+\,  _2F_1\left(\frac{1}{2},\frac{\nu}{2+2\nu};\frac{2+3\nu}{2+2\nu};\frac{Z^{2+\frac{2}{\nu}}}{Z_*^{2+\frac{2}{\nu}}}\right)\right)+{\tilde{\chi}}(u),~~~~~~~
\eea%
where $\chi$ and $\tilde{\chi}$ are arbitrary functions. To have better clarification of $\chi$ and $\tilde{\chi}$, let's equalize the above equations and use \eqref{solZ}, we get
\be%
\chi(v) = 2 \phi(v),~~~~~{\tilde{\chi}}(u) = 2 \phi(u).
\ee%
An appropriate choice of the arbitrary function, introduced in \eqref{solZ}, is $\phi(y)=y$. Our calculations in this paper are done by the same choice.  For more details see \cite {Ishii:2014paa}.  

\end{itemize}

\section{Numerical results}

\begin{figure}[ht]
\begin{center}
\includegraphics[width=66 mm]{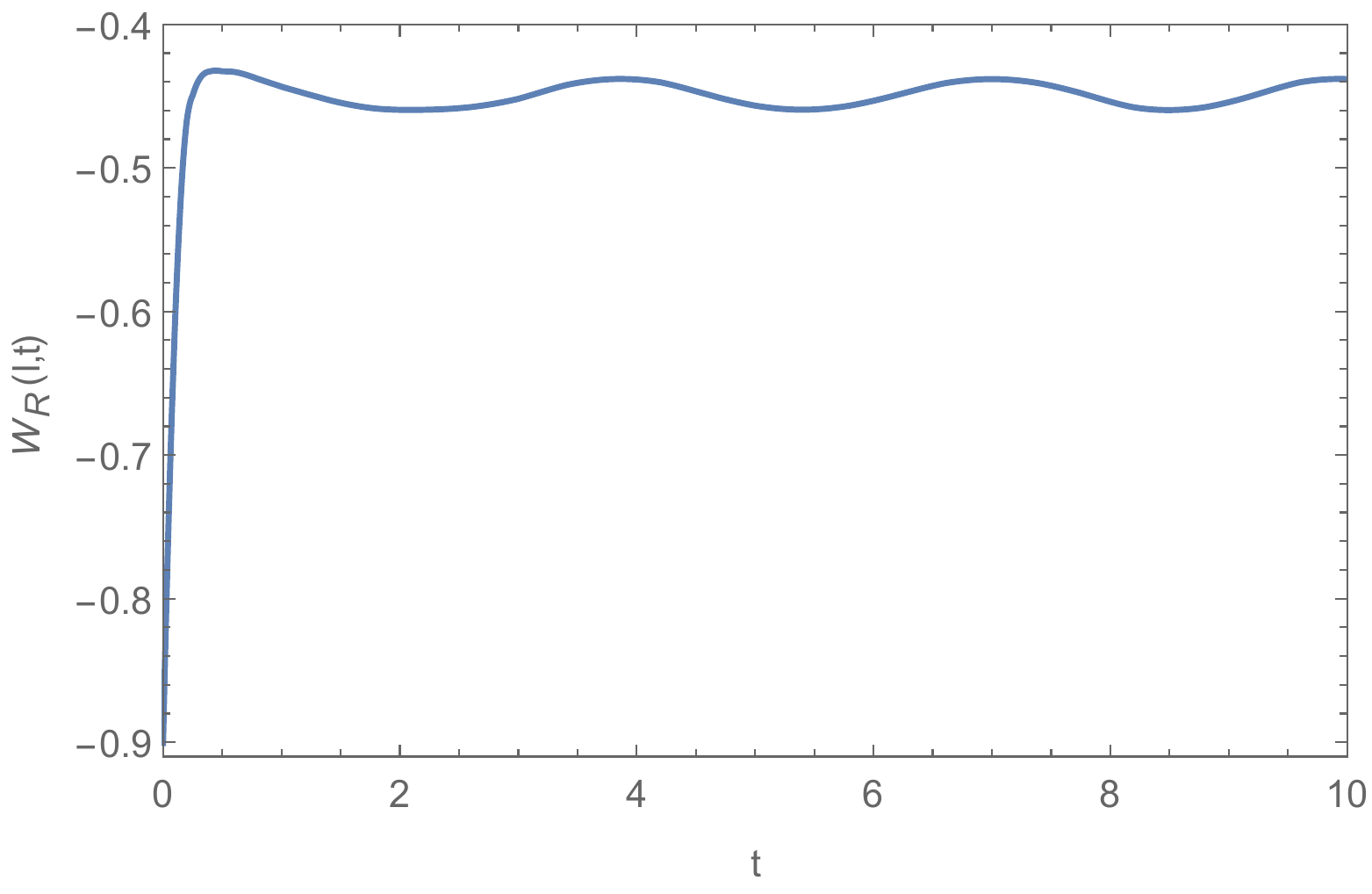}
\hspace{2 mm}
\includegraphics[width=66 mm]{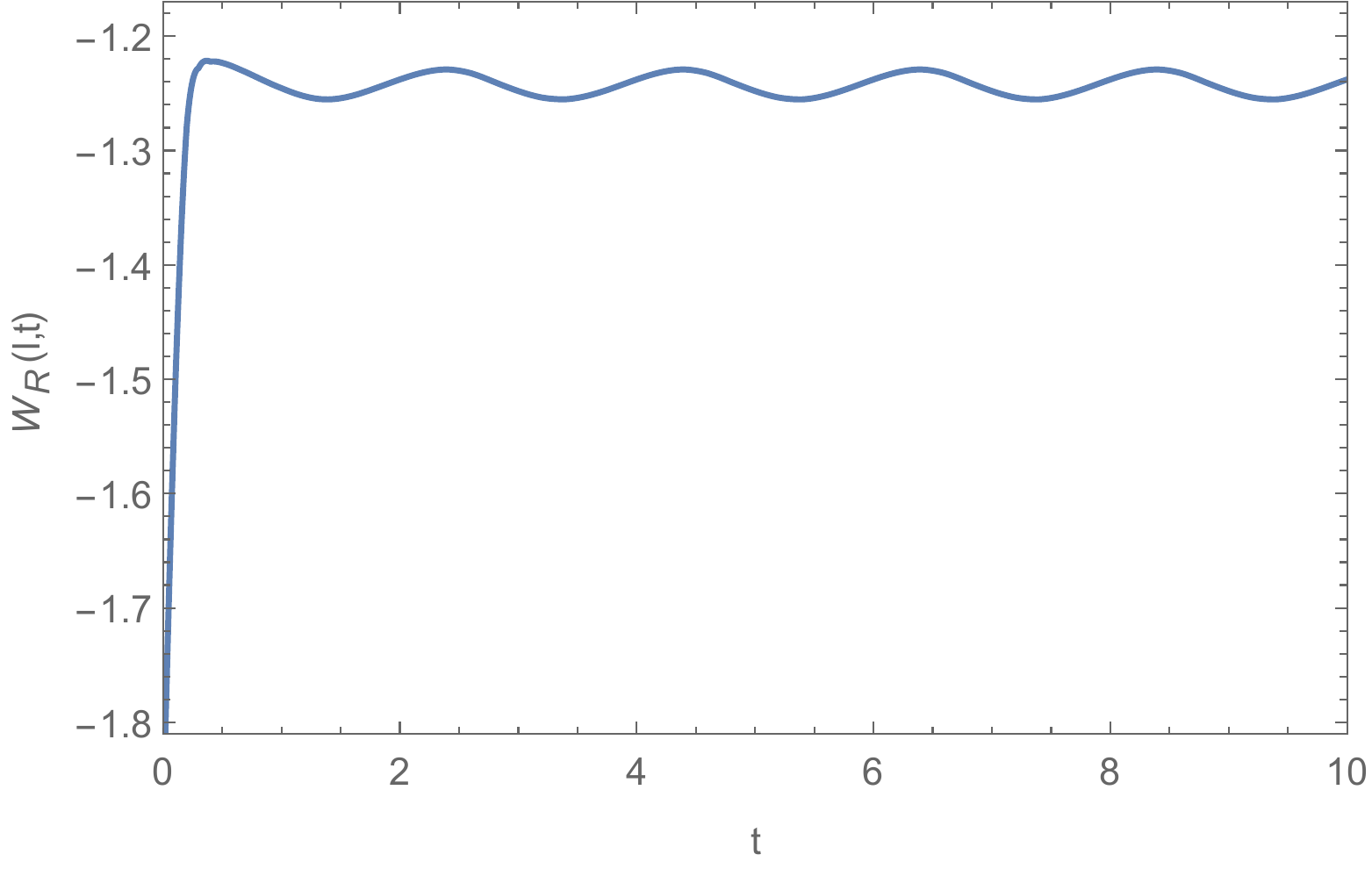}
\includegraphics[width=66 mm]{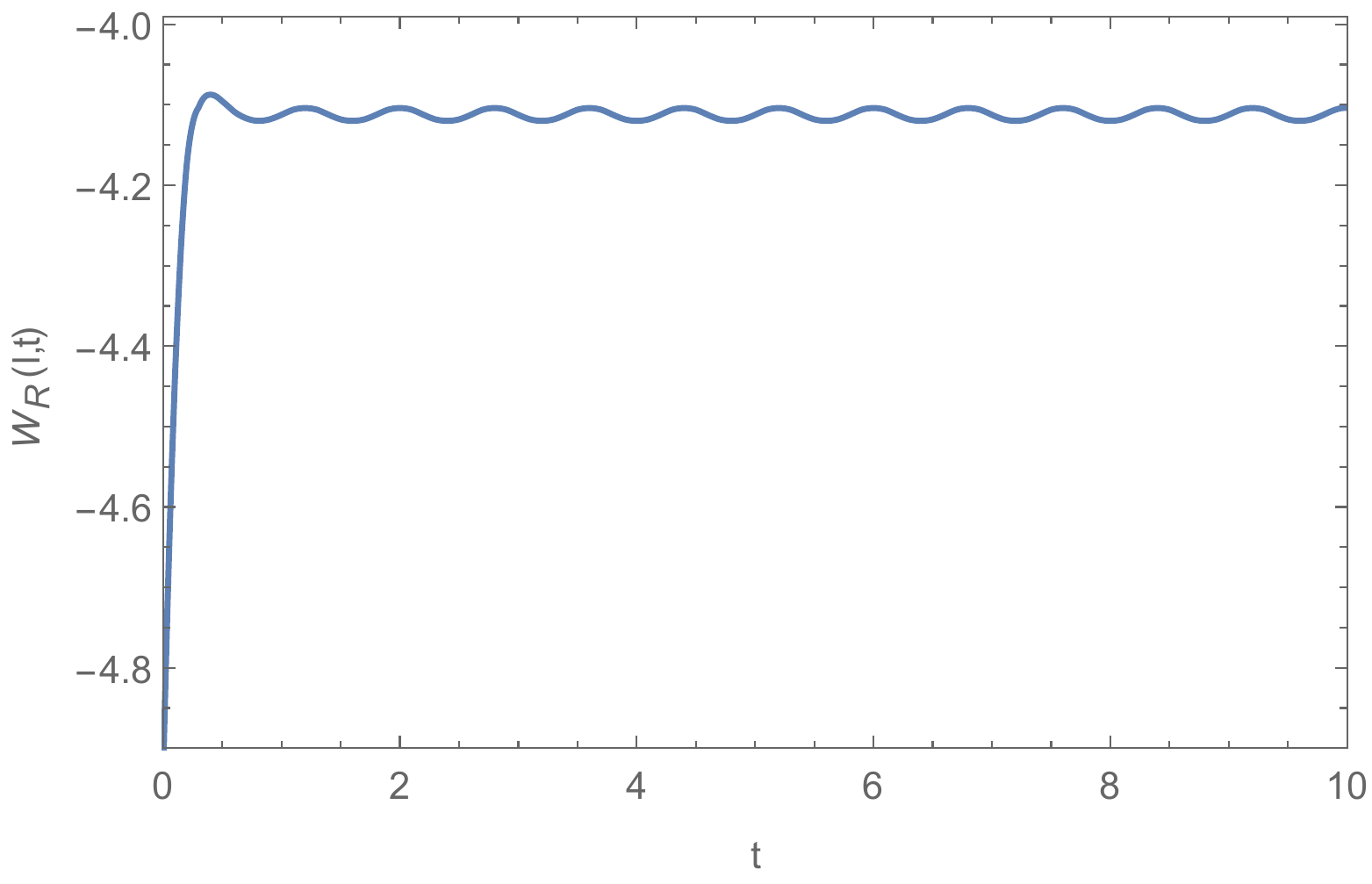}
\hspace{2 mm}
\includegraphics[width=66 mm]{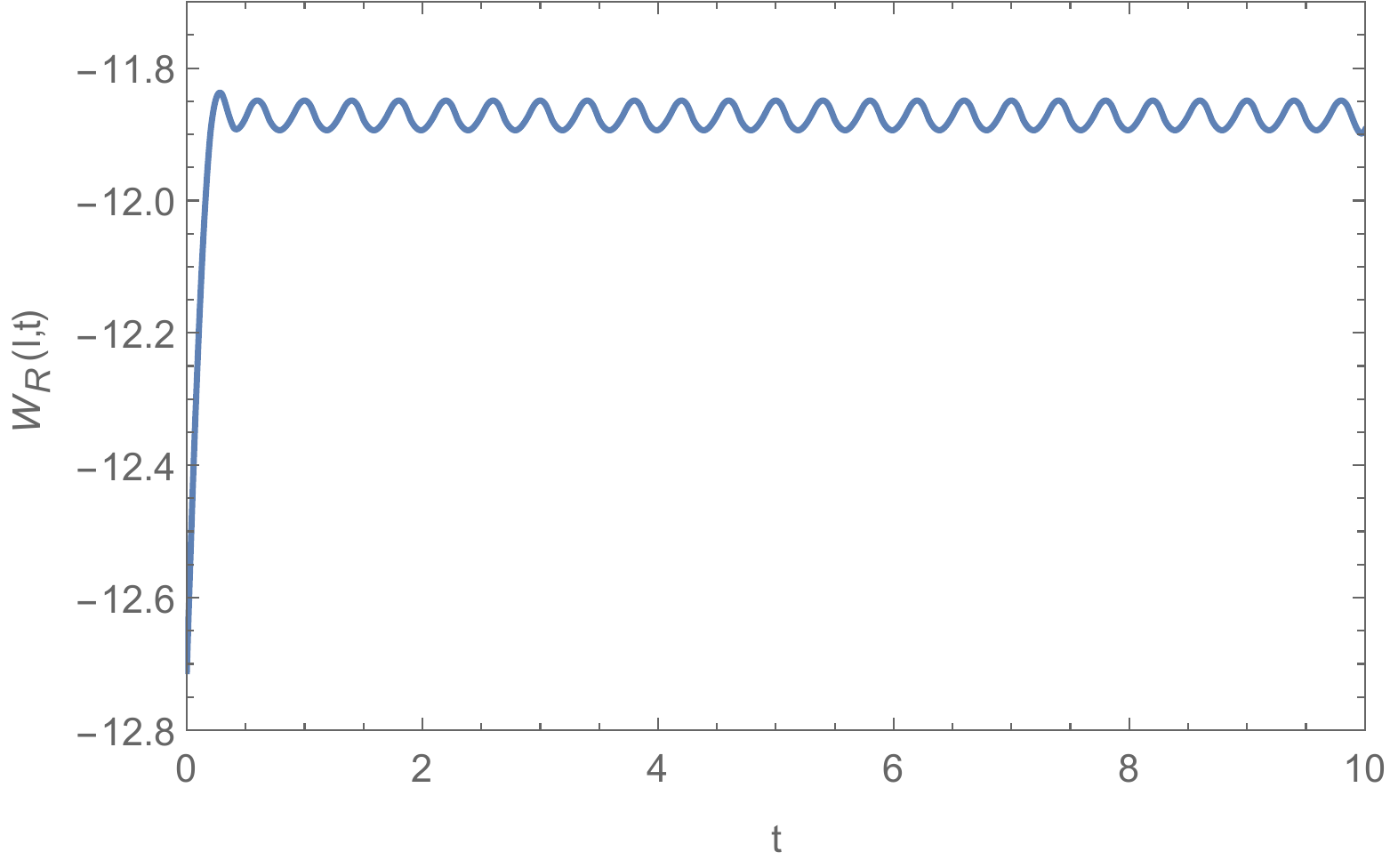}
\caption{ The expectation value of the Wilson loop in terms of boundary time for $\nu=1$ (top-left), $\nu=2$ (top-right), $\nu=3$ (bottom-left) and $\nu=4$ (bottom-right). For all figures we set $k=0.3$, $T_f=0.159$, and $l=1$. The static potential is $V(l)=-0.4488, -1.2422, -4.1120$ and $-11.8713$ coresponding to $\nu=1, 2, 3$ and 4, respectively. \label{veffect1}
}
\end{center}
\end{figure}%
\begin{figure}[ht]
\begin{center}
\includegraphics[width=66 mm]{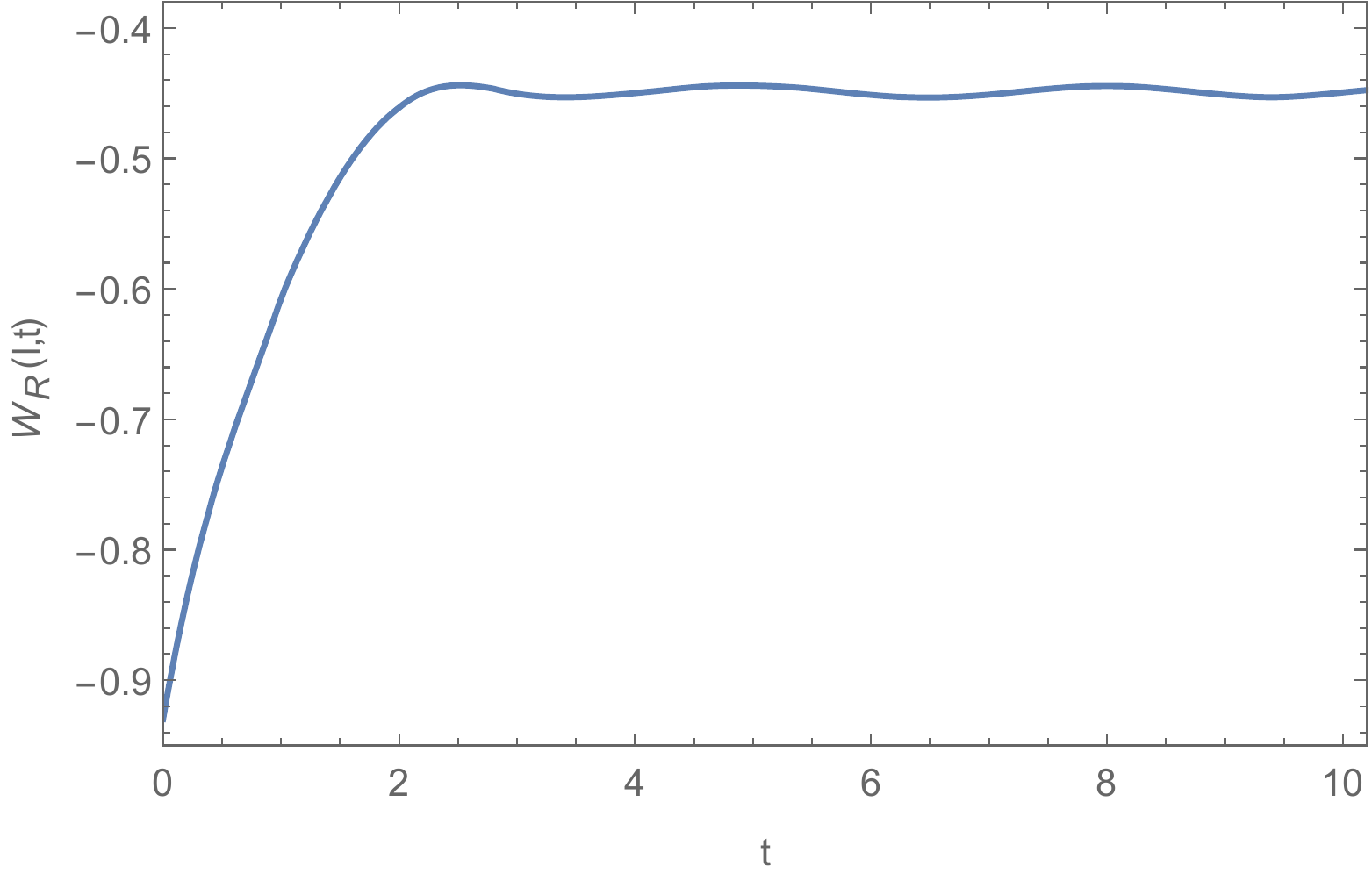}
\hspace{2 mm}
\includegraphics[width=66 mm]{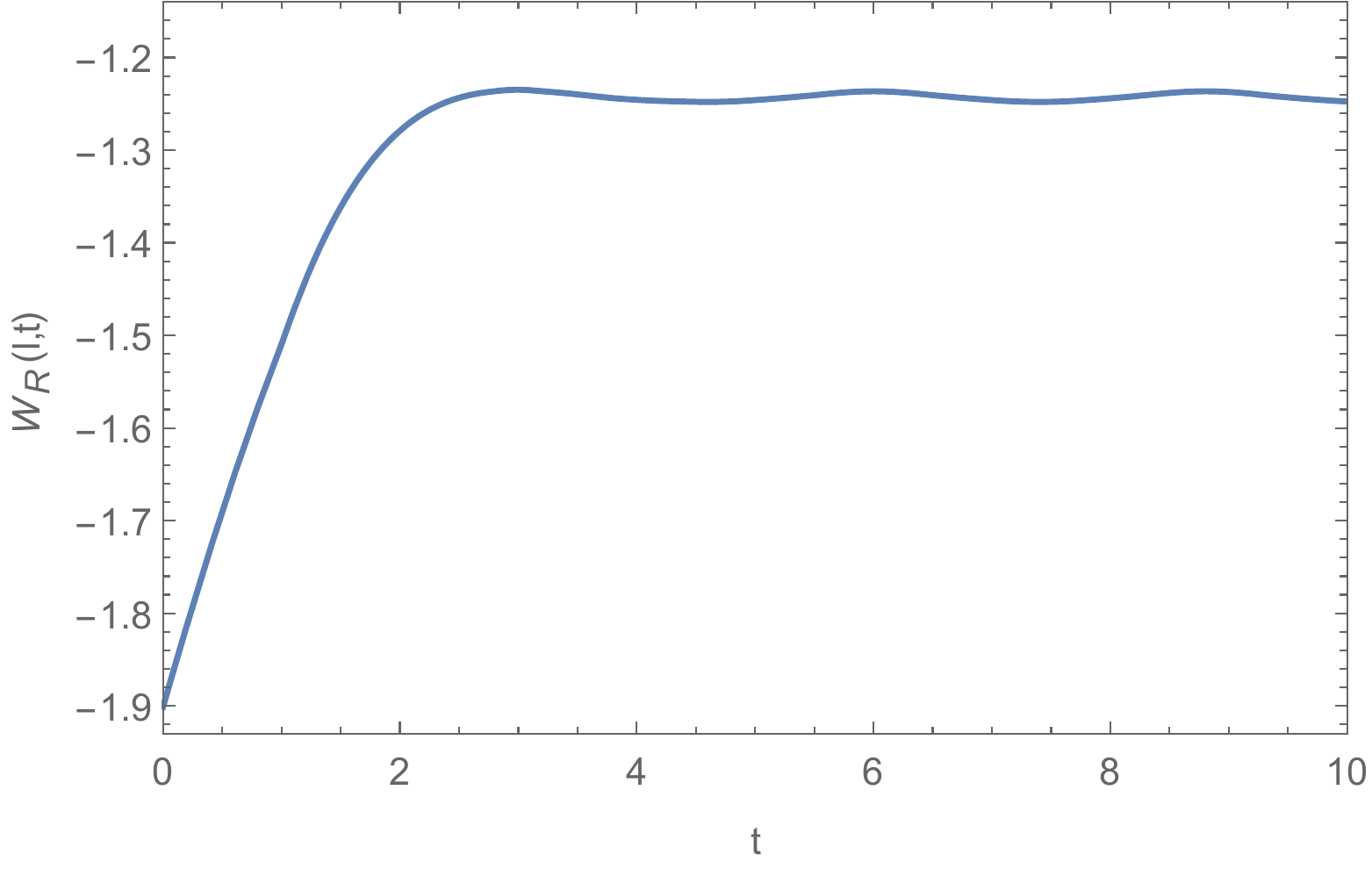}
\includegraphics[width=66 mm]{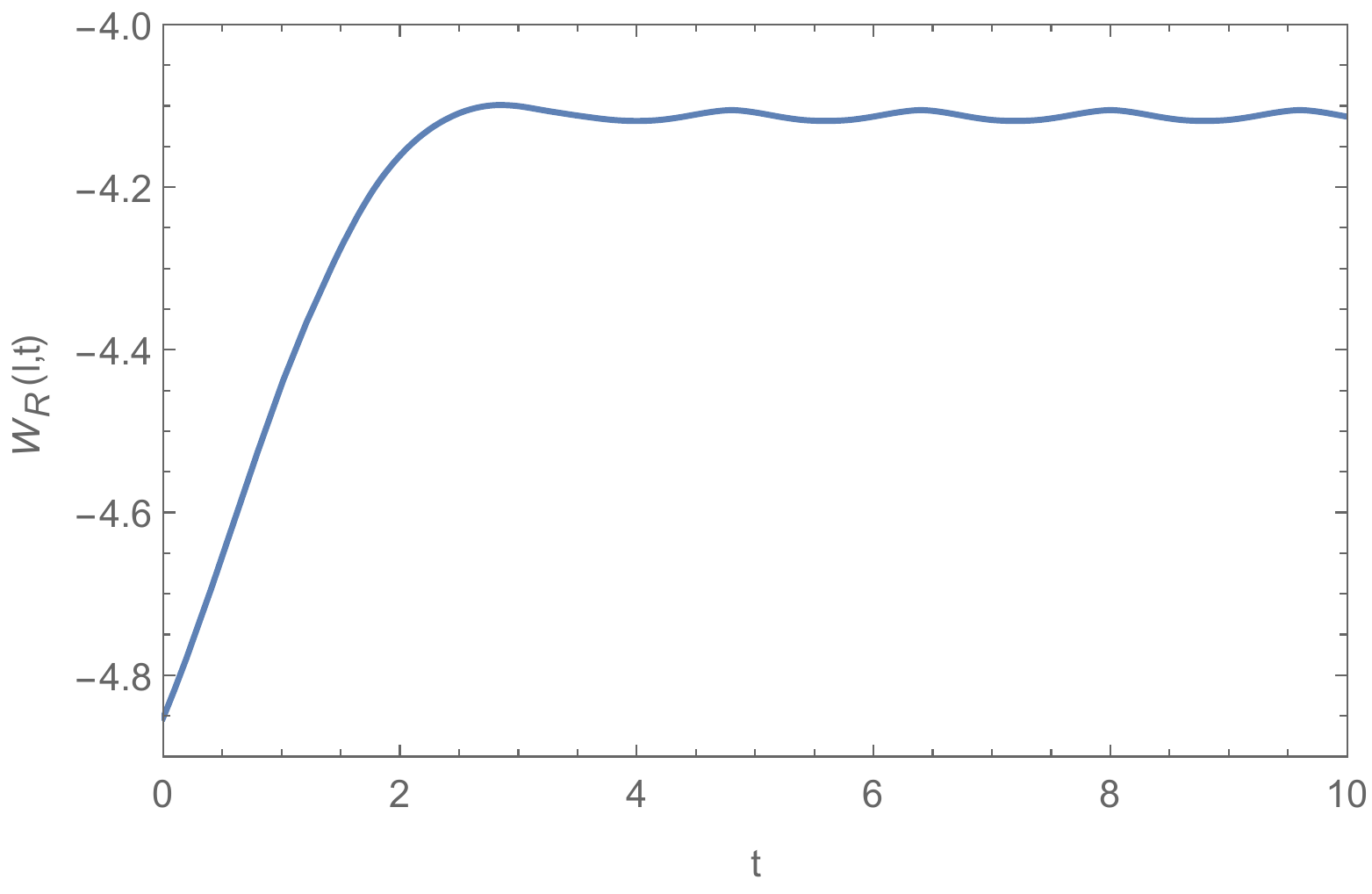}
\hspace{2 mm}
\includegraphics[width=66 mm]{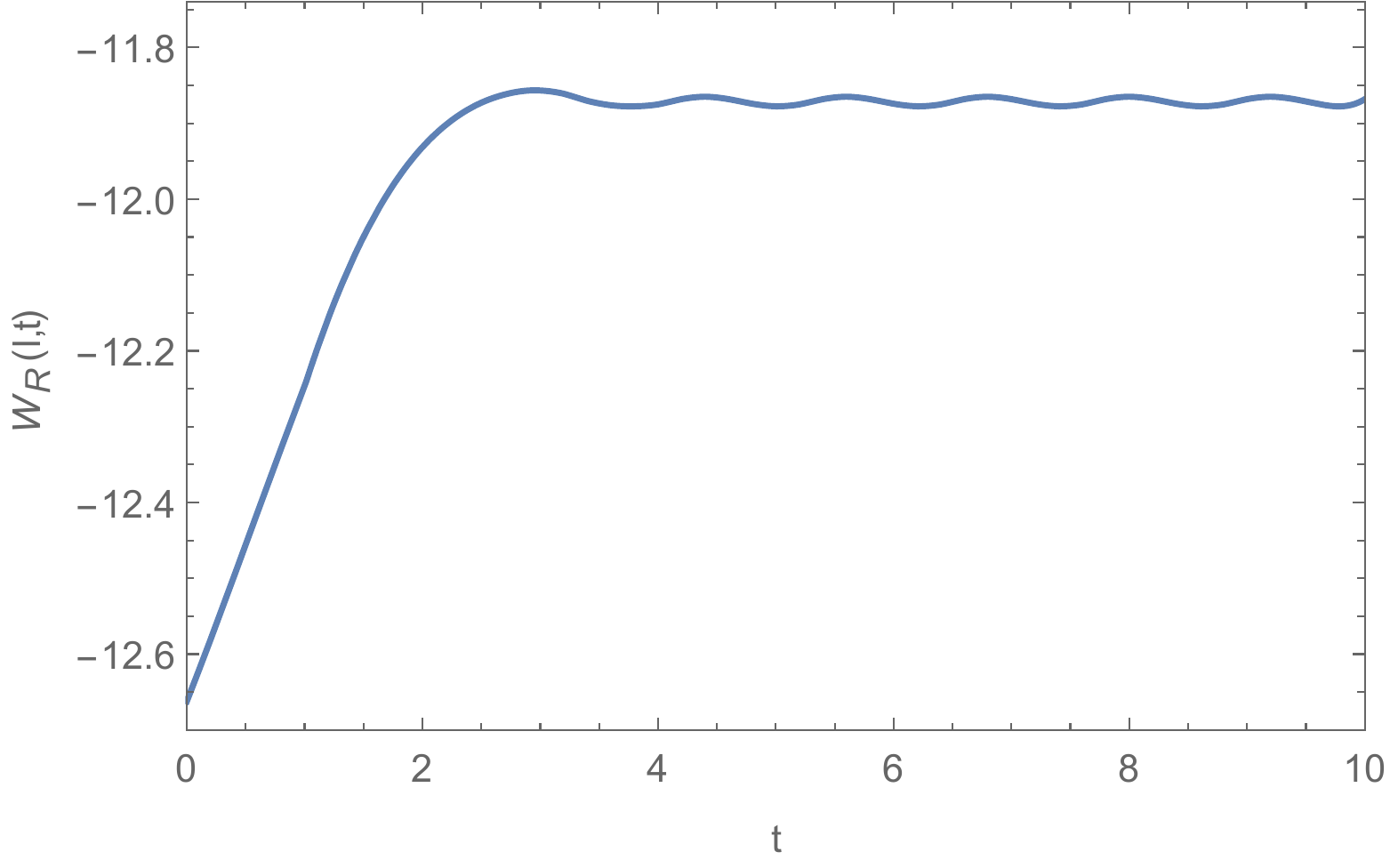}
\caption{The expectation value of the Wilson loop in terms of boundary time for $\nu=1$ (top-left), $\nu=2$ (top-right), $\nu=3$ (bottom-left) and $\nu=4$ (bottom-right). For all figures we set $k=3$, $T_f=0.159$, and $l=1$. The static potential is $V(l)=-0.4488, -1.2422, -4.1120$ and $-11.8713$ coresponding to $\nu=1, 2, 3$ and 4, respectively. \label{veffect2} 
}
\end{center}
\end{figure}%

Based on the gauge-gravity duality, classical string in the anisotropic background is dual to a quark-anti-quark bound state in the anisotropic gauge theory. When the boundary time is smaller than zero, that is $t<0$, the meson is stable and in its ground state. As the energy injection is started, or equivalently the temperature is raised on the gauge theory side, the shape of the string changes time-dependently. In fact, during the energy injection, the turning point of the string goes closer to the horizon in the background. Our numerical results show that the string oscillates about the static string solution corresponding to the final temperature of the energy injection. In the gauge theory, this is the reason why expectation value of the Wilson loop oscillates about static potential by which we mean the potential of the bound state in the anisotropic plasma with finite temperature $T_f$. These oscillations indicate that bound state is excited by energy injection. This result is in agreement with the similar one reported in \cite{Ali-Akbari:2015ooa}. Note that since there is no energy dissipation, the excited meson is stable.

\begin{figure}[ht]
\begin{center}
\includegraphics[width=66 mm]{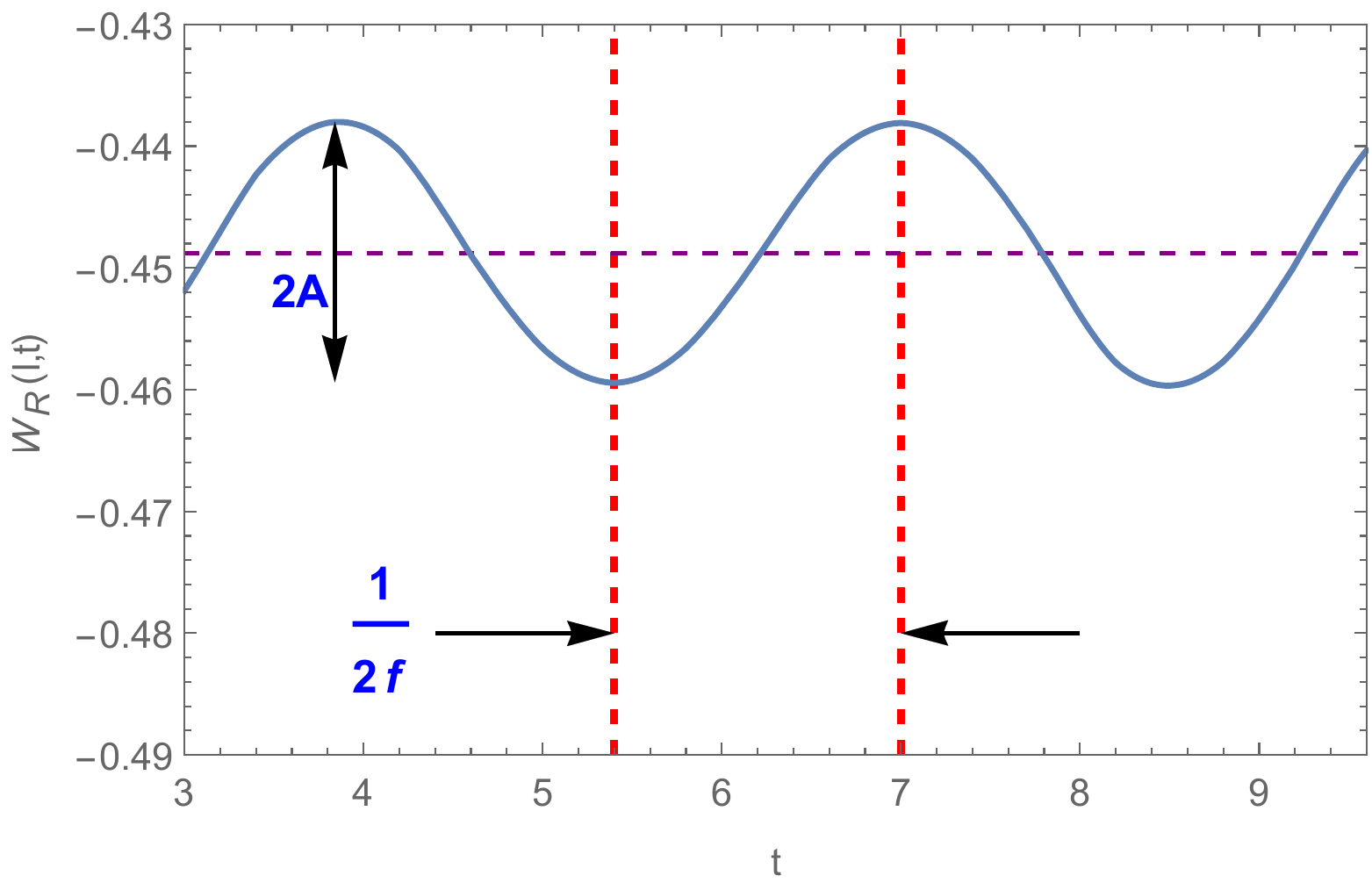}
\hspace{2 mm}
\includegraphics[width=66 mm]{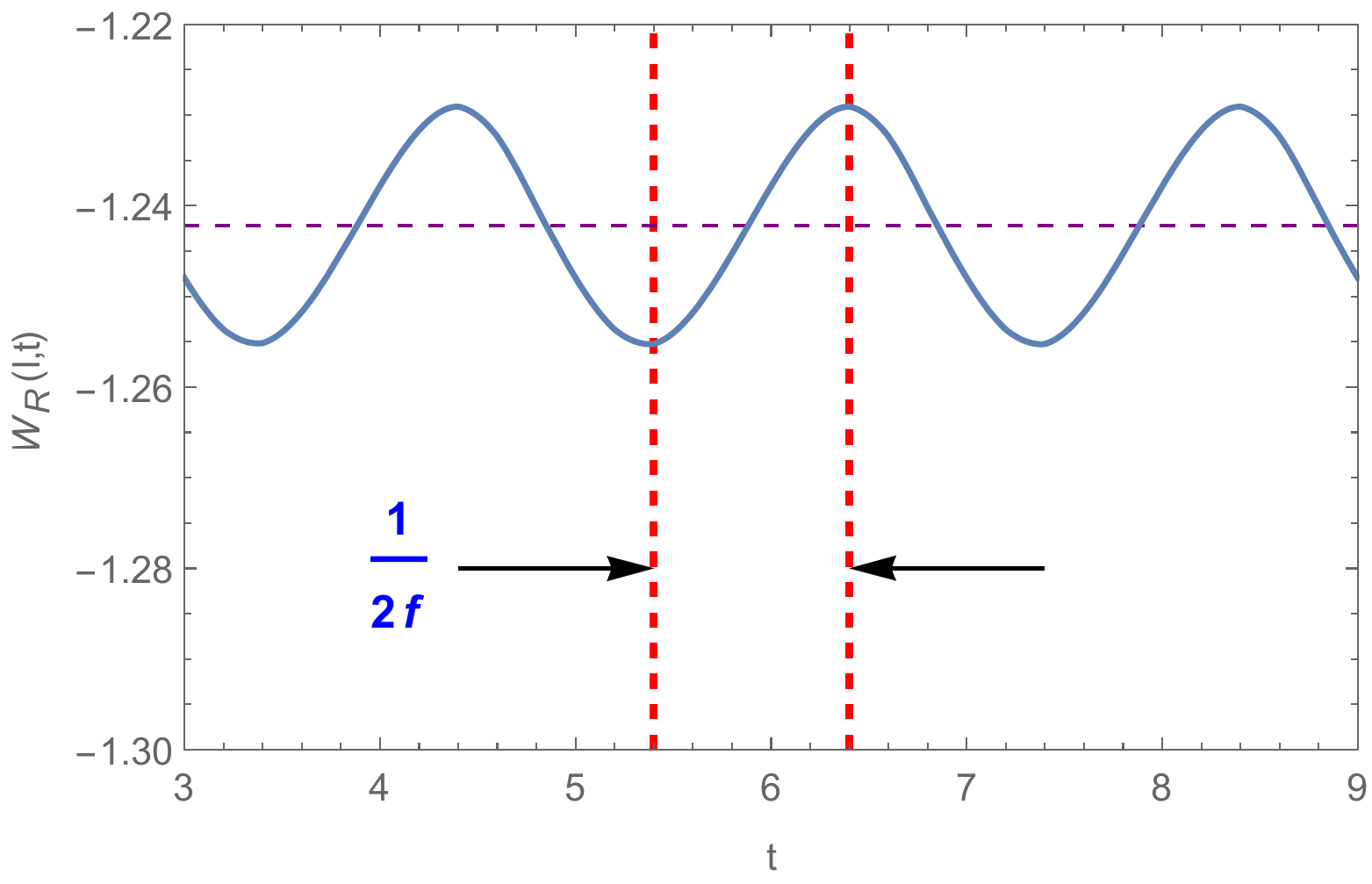}
\includegraphics[width=66 mm]{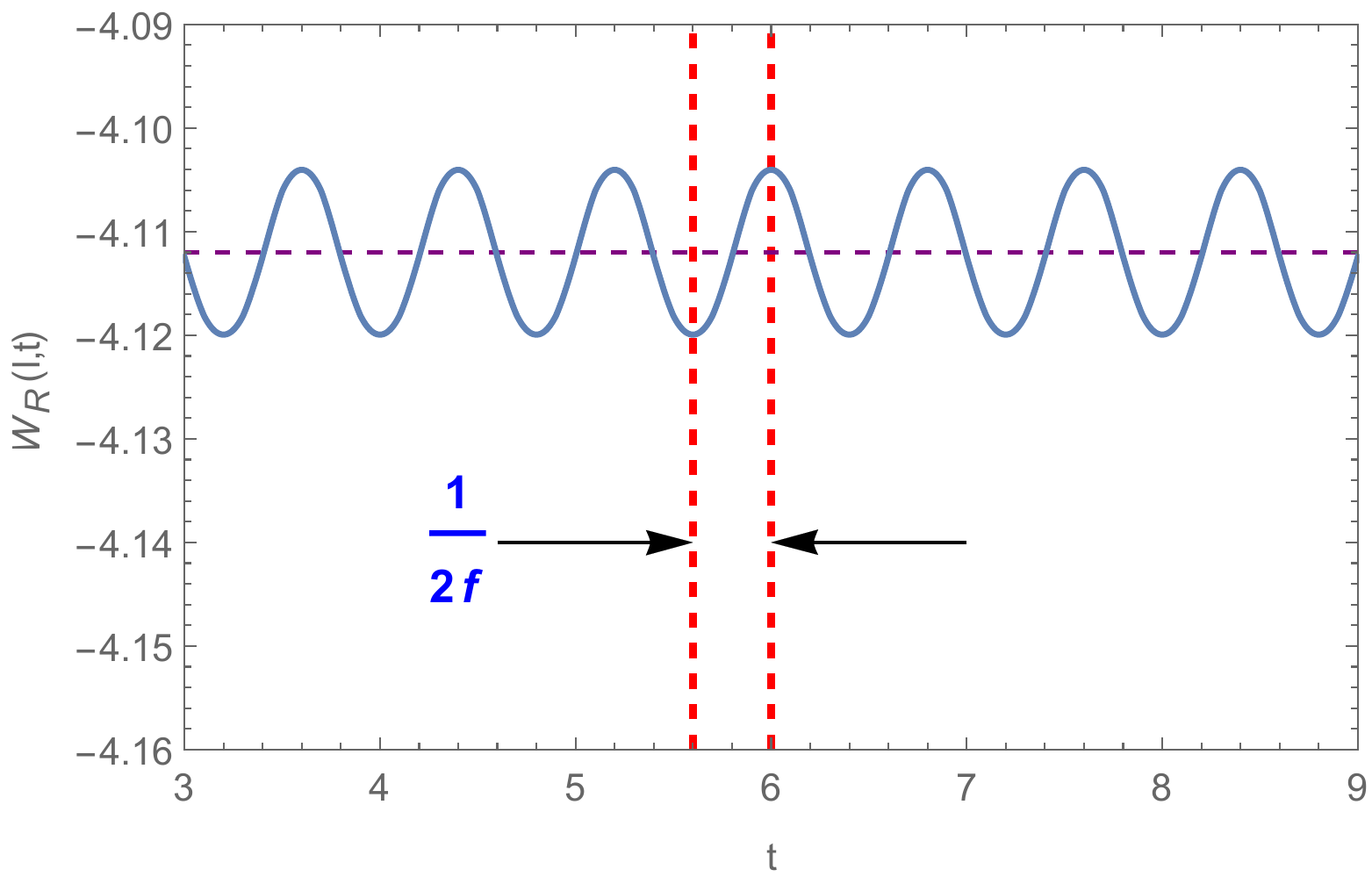}
\hspace{2 mm}
\includegraphics[width=66 mm]{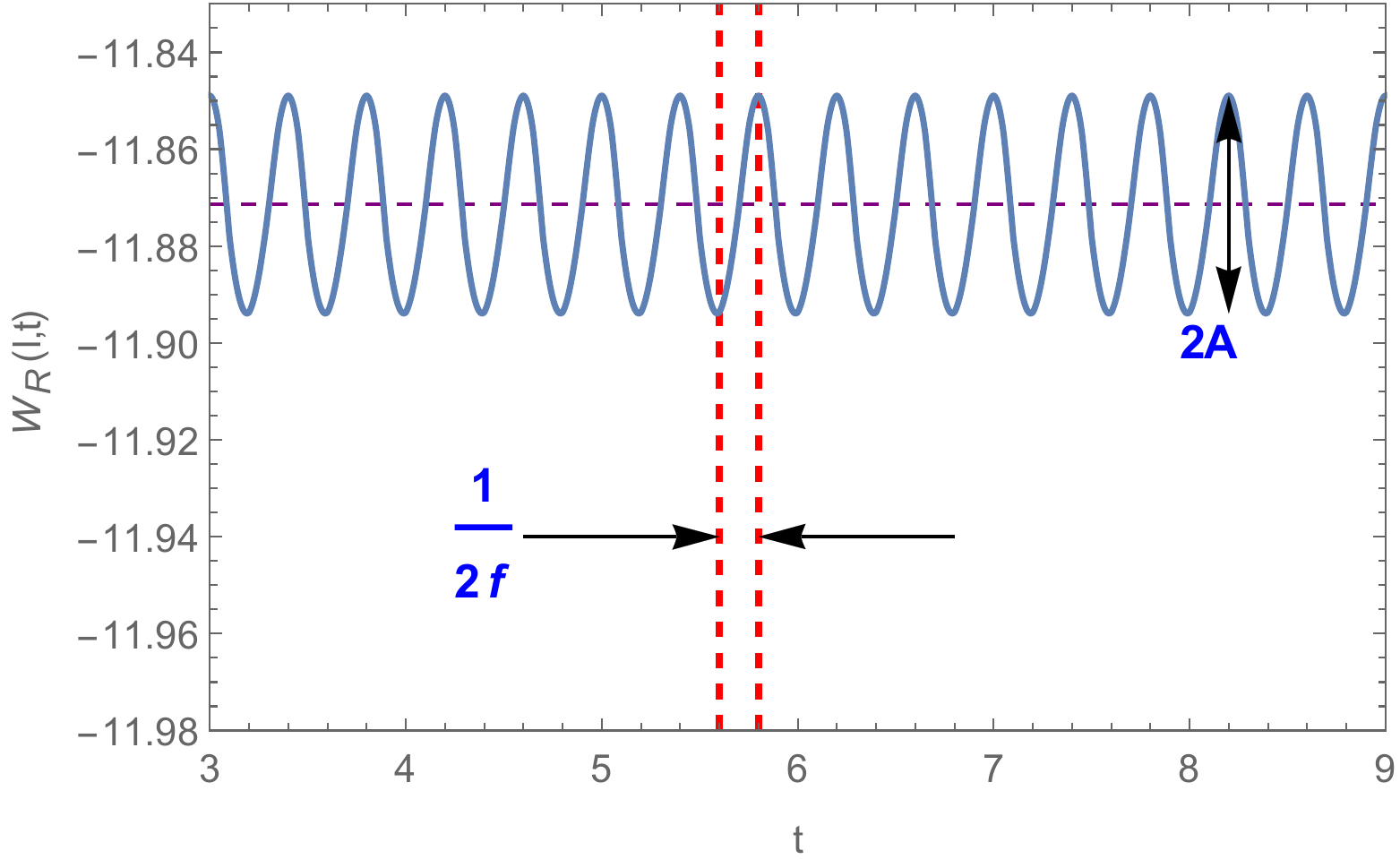}
\caption{The expectation value of the Wilson loop in terms of boundary time for $\nu=1$ (top-left), $\nu=2$ (top-right), $\nu=3$ (bottom-left) and $\nu=4$ (bottom-right). For all figures we set $k=0.3$, $T_f=0.159$, and $l=1$. The static potential is $V(l)=-0.4488, -1.2422, -4.1120$ and $-11.8713$ corresponding to $\nu=1, 2, 3$ and 4, respectively. The dashed purple line shows the static potential in each case. For $\nu=1$ and 4, we have $2A=0.0213$ and $0.0447$, respectively.
\label{shape3}}
\end{center}
\end{figure}

Understanding how anisotropy parameter $\nu$ affects on the time evolution of the expectation value of the Wilson loop is an interesting issue to investigate. In figure \ref{veffect1}, ${\cal{W}}_R(l,t)$ has been plotted for various anisotropy parameters at fixed values of transition time $k$, final temperature $T_f$ and distance $l$. It is clearly seen that the excited bound state is different for each anisotropy parameter. More precisely, the larger anisotropy parameter, the larger frequency. Furthermore, independent of the anisotropy parameter, the time-dependent expectation value starts oscillating around the negative equilibrium value of the static potential almost at the same time, i.e. $t \simeq 0.30$. In figure \ref{shape3}, we show that the amplitude of the oscillation increases for larger $\nu$, too.

Apart form the transition time $k$, other variables are the same in the figures \ref{veffect1} and \ref{veffect2}. Evidently, the results extracted from the figure \ref{veffect2} is similar to the case in the figure \ref{veffect1}. The only difference is that the bound state oscillate with a lower frequency and amplitude in this case.

\begin{table}[ht]
\caption{The oscillation frequency of figure \ref{shape3} for various anisotropy parameters}
\vspace{3 mm}
\centering
\begin{tabular}{c c c c c}
\hline\hline
~~$ \nu$ ~~   &   ~~ $1$ ~~ & ~~ $2$ ~~ & ~~ $3$ ~~ & ~~ $4$ ~~\\[0.5ex]
\hline
$f$ & 0.31 & 0.50 & 1.25 & 2.50\\
\hline
\end{tabular}\\[1ex]
\label{ahmad}
\end{table}

\begin{figure}[ht]
\begin{center}
\includegraphics[width=66 mm]{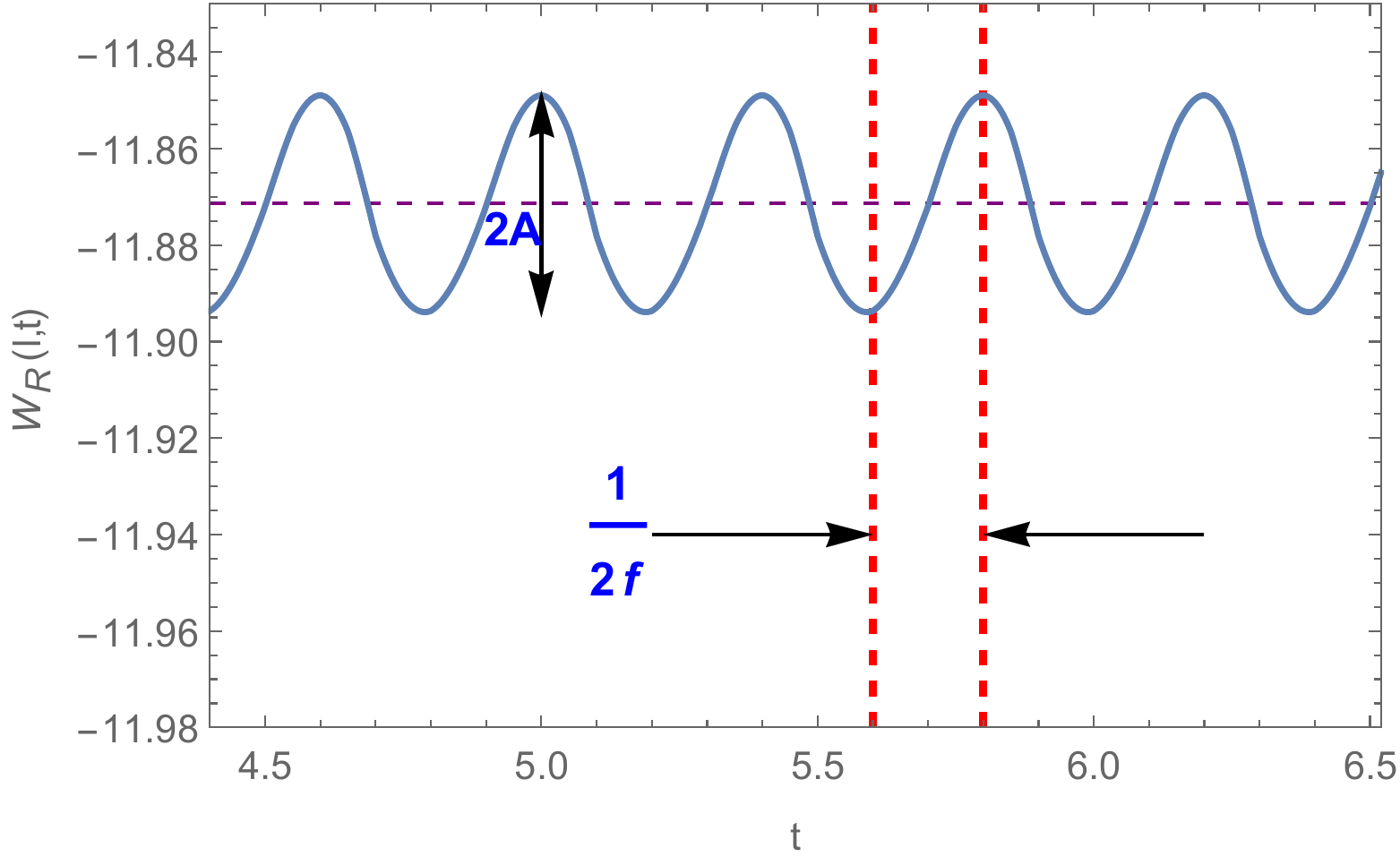}
\hspace{2 mm}
\includegraphics[width=66 mm]{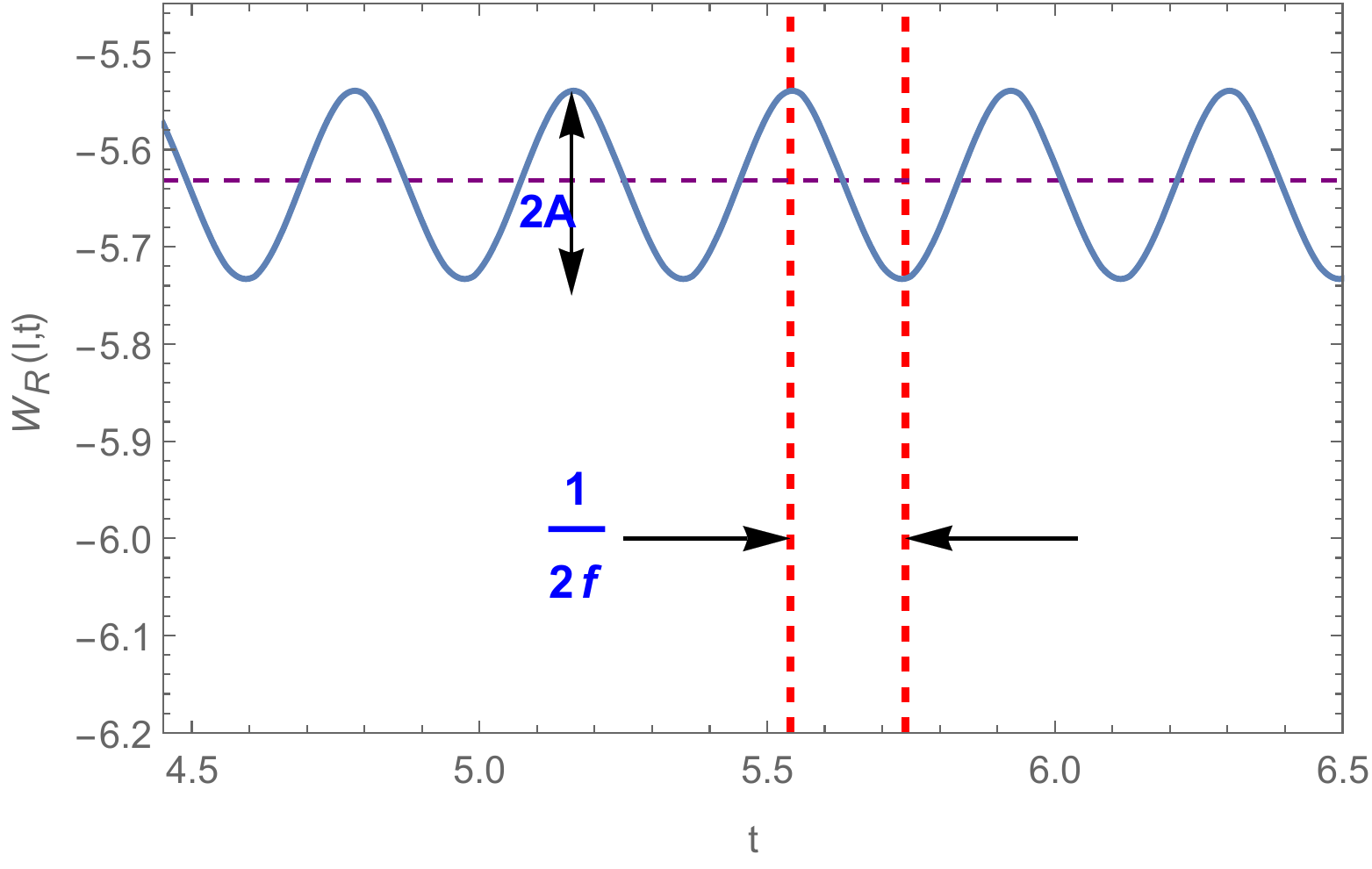}
\caption{The expectation value of the Wilson loop in terms of boundary time for $\nu=4$, $k=0.3$ and $l=1$. The temperature for the graph left (right) are $0.159$ ($0.796$) and oscillation frequency is $0.19$ for both cases. The static potentials are -11.8713 and -5.6318. The amplitude of oscillation is $2A=0.0447$ for the left panel and $2A=0.1869$ for right one.
\label{figure34}}
\end{center}
\end{figure}%

\begin{figure}[ht]
\begin{center}
\includegraphics[width=66 mm]{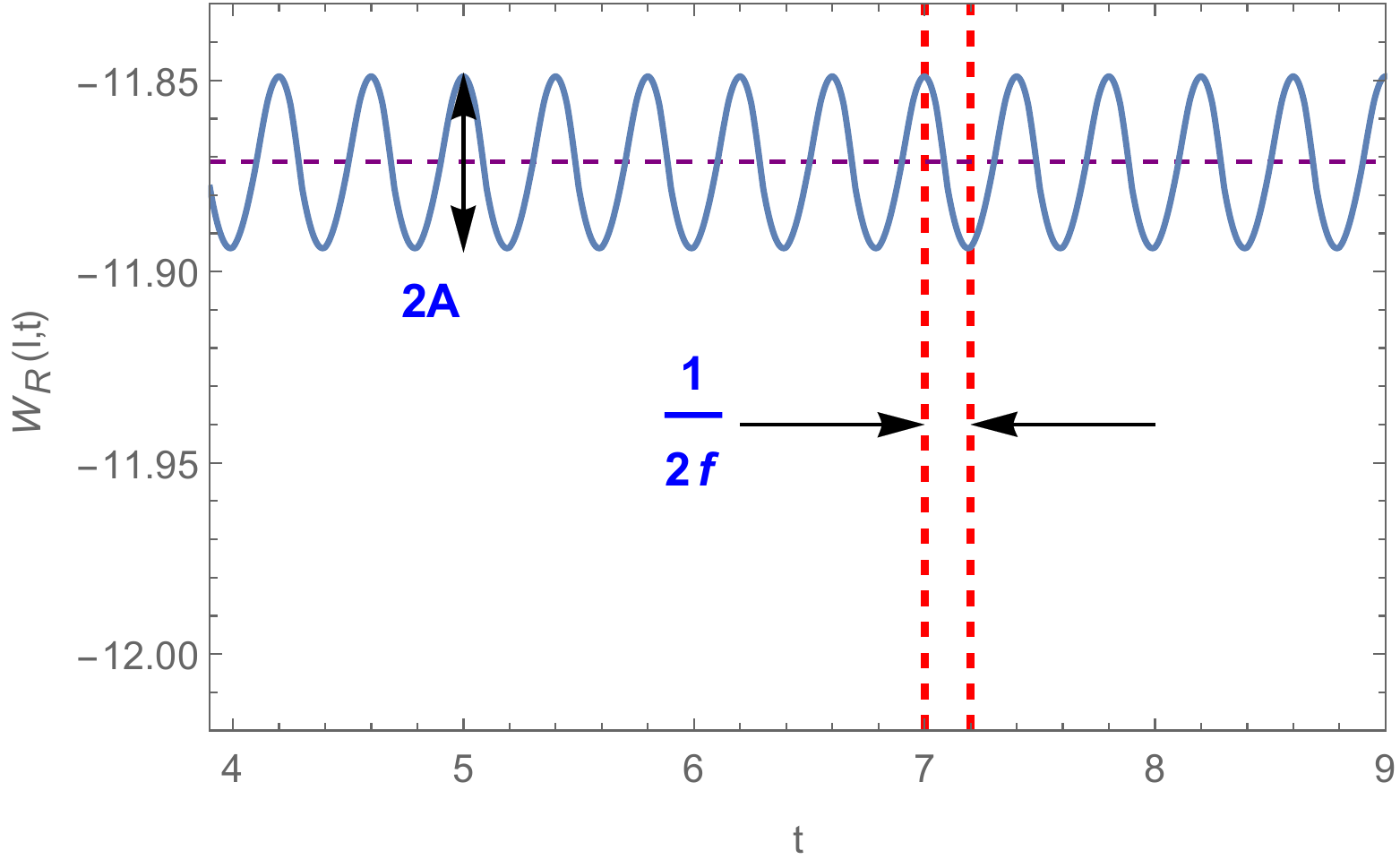}
\hspace{2 mm}
\includegraphics[width=66 mm]{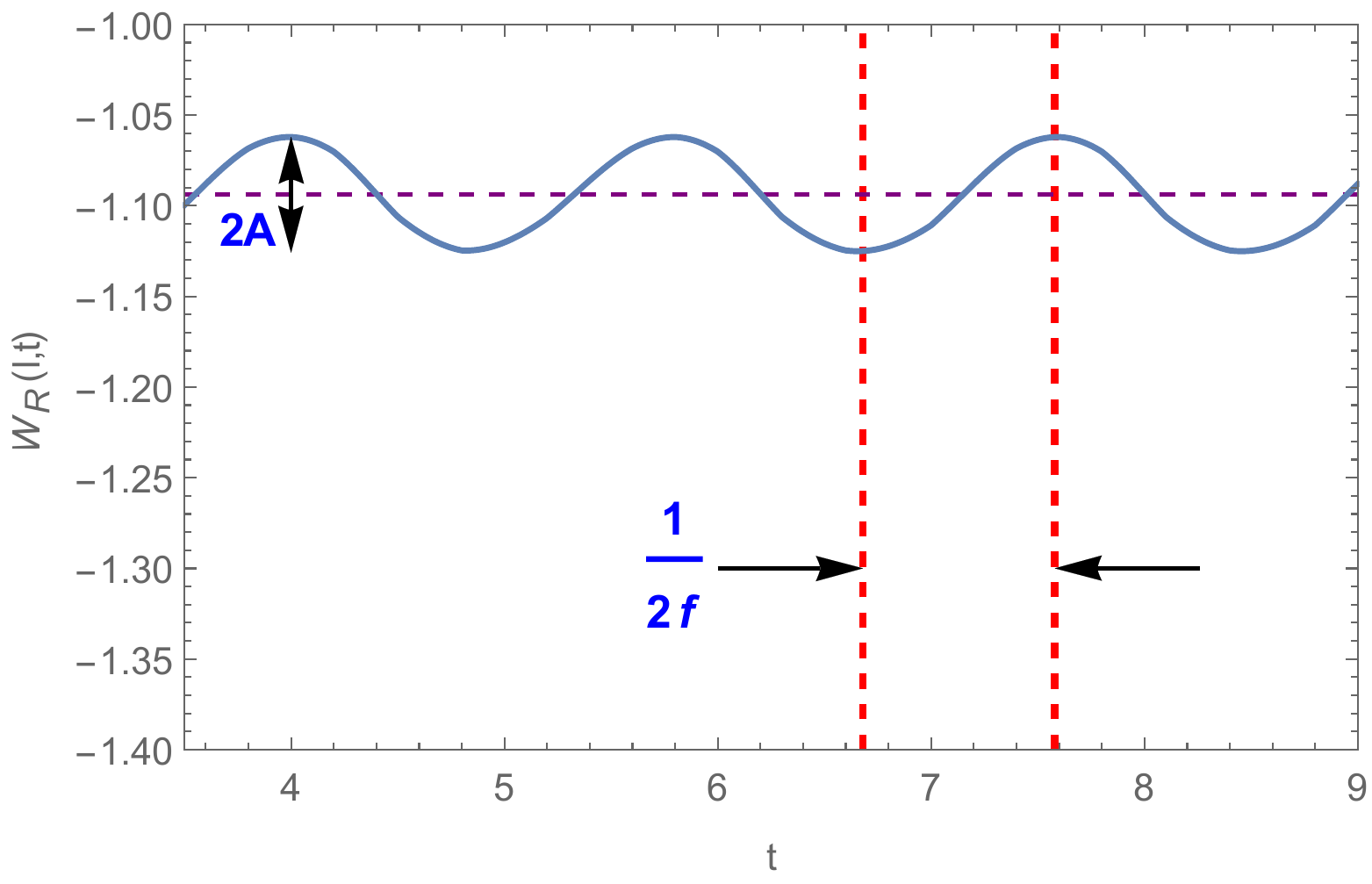}
\caption{The expectation value of the Wilson loop in terms of boundary time for $\nu=4$, $k=0.3$, $T=0.159$ and $l=1 (l=1.5)$ for the left (right) figure. The oscillation frequency is $2.5 (0.6)$ for the left (right) figure. The static potentials are -11.8713 and -1.0937. The amplitude of oscillation is $2A=0.0447$ ($0.0631$) for the left (right) panel.
\label{figurefinal}}
\end{center}
\end{figure}%

In order to have better estimate of the dependence of the frequency on the anisotropy parameter, we have plotted the expectation of the Wilson loop in term of the boundary time in the region of $t=3-9$ in figure \ref{shape3}. As it is clearly seen from this figure and confirmed by table \ref{ahmad}, the oscillation frequency substantially increases when the anisotropy parameter is raised in the plasma. It is important to notice that although the frequency of the excited bound state is larger, it is deeply bounded due to the anisotropy in the system. 

We would also like to investigate the effect of the temperature on the oscillation frequency. To do so, we plot the expectation value of the Wilson loop in terms of boundary time for two different final temperatures in figure \ref{figure34} for fixed values of anisotrpy parameter, i.e. $\nu=4$, and distance $l$. The temperature in the left graph is less than the right one. Interestingly, the frequency is the same for both cases. It means that the oscillation frequency is independent of final temperature. However, this figure indicates that the amplitude of the oscillation depends on the final temperature and they increase together. The same behaviour is also observed in case with $k=3$.

Finally in the figure \ref{figurefinal}, we have plotted the time evolution of the expectation value of the Wilson loop in terms of boundary time for two different values of distance $l$. At larger distances, the amplitude of the oscillation increases while the oscillation frequency decreases. Therefore, the oscillation characteristics depend on the distance $l$, too.

To summarize, we find that the oscillation frequency is independent of time and final temperature, i.e. $f(l,k,\nu)$. However, the amplitude of oscillation depends on the all parameters in the theory, that is $A(l,T_f,k,\nu)$. Notice that neither the frequency nor the amplitude of the oscillation does not change with the time since the bound state is living in the plasma without dissipation. From our results, one can conclude that the amplitude of the oscillation increases when each parameter of the problem at hand raises.
It is then interesting to compare our results with harmonic oscillator model. If we consider $M(V)$, corresponding to the time-dependent temperature in the gauge theory, as an external force, the (average) energy of the bound state increases due to energy injection. This enhancement is more substantial in the presence of the anisotropy as well as at higher final temperatures.

\section*{Acknowledgment}
M. A. would like to thank School of Physics of Institute
for research in fundamental sciences (IPM) for the research facilities and
environment.

\end{document}